# Convergence of 5G with Internet of Things for Enhanced Privacy

Amreen Batool[1], Baoshan Sun[1], Ali Saleem[2], and Jawad Ali[3(✉)]

[1] School of Computer Science and Software Engineering, Tiangong University, Tianjin, China
{Amreen,sunbaoshan}@tiangong.edu.cn
[2] Department of Computer Science, COMSATS University Islamabad, Lahore, Pakistan
Ddp-sp15-bse-115@ciitlahore.edu.pk
[3] Malaysian Institute of Information Technology,
Universiti Kuala Lumpur, Kuala Lumpur, Malaysia
jawad.ali@s.unikl.edu.my

**Abstract.** In this paper, we address the issue of privacy in 5th generation (5G) driven Internet of Things (IoT) and related technologies while presenting a comparison with previous technologies for communication and unaddressed issues in 5G. Initially, an overview of 5G driven IoT is presented with details about both technologies and eventually leading to problems that 5th generation will face. Details about 5G are also presented while comparing them with previous technologies. The architecture of 5G is presented hence explaining layers of 5G and technologies like SDN, NFV and cloud computing that compose these layers. The architecture for 5g based IoT is also presented for providing visual understanding as well as explained based on how this addresses the issues present in 4G. Privacy is highlighted in 5G driven IoT while providing details about how SDN, NFV and cloud computing helps in elimination of this issue. The issues presented will be compared with 4G based IoT and solutions are provided about mitigation of these issues particularly bandwidth and security. Moreover, techniques used by 4G and 5G technologies for handling the issues of privacy in IoT are presented in a nutshell as a table. Paper also presents a detailed overview of technologies making 5G possible meanwhile giving an explanation about how these technologies resolve privacy issues in 5G.

## 1 Introduction

Internet of Things (IoT) can be considered as a network of objects, animals or machines that transfer data without involvement of human or computer interaction. The statement above defines it as a more open and independent system that benefits in terms of content transfer and interaction, but it also possesses challenges like the need of more bandwidth for transfer of massive content as well as low latency and security. IoT is an open network in terms of devices connected as any node is malicious to another node which rises trust and privacy issue in IoT [1–3]. Currently, IoT is deployed on current technology 4G that lacks the inability to address the challenges of IoT specifically bandwidth and privacy [4].

Scarcity of resources, as well as security and privacy issues, have raised the need for a more versatile and secure technology that make sure the integrity of data as it is transferred over the channel. 5G is on the verge of deployment making massive IoT and tactile internet technologies possible by addressing data rate and latency issues. Meanwhile providing enhanced security and privacy for these technologies.

Massive IoT implementation over 5G will be possible with software-defined networking that will not only enhance the speed but also will address many privacy issues faced by the current IoT network [5]. Hence in an environment where devices are connected wirelessly, it is of critical importance that devices communication is secure and nodes privacy is not compromised. 5G will provide a higher data rate of up to 10 Gbps and with more bandwidth and ability to choose among various service providers for QoS [6].

The ability to choose among several service provider will raise authentication and data integrity issues in the IoT environment. Moreover, data confidentiality, privacy availability and access control will also be an issue in open networks [7]. Furthermore, as 5G will be deployed over the current 4G IP base network it will by default inherit the problems specific to IP bases communication. Hence, to ensure the security and privacy in 5G for IoT and related technologies will certainly be a problem [8].

5G will not only benefit the IoT of things but also cloud, social computing, and cognitive radio technology as well. Moreover, it is designed by keeping in mind the needs for virtualization that will enable novice services that will also raise security and privacy issues. IoT has revolutionized the communication among nodes by the distribution of processing and storage power among mobile devices [9, 10]. IoT generic layered architecture constitutes of three layers that enable the IoT services provided to applications. This distribution of processing power and storage among multiple issues raise much security, privacy and trust issues in IoT enables devices [11]. Thus, a key step towards the resolution of these problems is IoT middleware that is a logical layer.

Meanwhile, Integration of 5G with IoT provides an SOA approach for IoT that will enable the decomposition of IoT system and provide a service-oriented approach for privacy threats in IoT architecture [12].

Moreover, our paper is portioned in 3 different sections. The first section is about the introduction of the topic as well a brief overview of what has already been done in the concerned field. The second section discusses the evolution of technologies from the first generation to 5G as well as a comparison table about what privacy techniques used in all these technologies to mitigate privacy and security concerns. The third section discusses the architecture of 5G along with its different layers and what measurements can be taken to enhance the technology that is on the verge of deployment. Forth section discusses the integration of 5G with IoT and various other domains like mobile computing and cloud computing and how the integration of these technologies will control the privacy risks while providing delay requirement of less than 10 s for real-time communication.

## 2   1G to 5G Transition

Wireless network communication architecture starting its journey from a 1G analog system to an era of 4.5G more flexible and secure IP-based network designed specifically for mobile networks has been successfully implemented. Although a couple of years back 4.5G was a secure enough system for IoT, cloud communication and related technologies [13, 14]. On the contrary the expectations proved to be wrong; with the arrival of the internet of everything (IOE) and tactile internet the network required should have ample bandwidth along with greater data rate to achieve the lesser delay requirement for real-time communication [15]. Although the hype created about the deployment of 5G is believed to not fulfil many expectations of the industry. Starting our comparison from a 1G analog system that was developed for only voice communication. Considering the security concerns in first-generation communication system 2G was introduced as a digital voice and SMS system that provided enhanced coverage, as well as the better data rate of 64kbps for faster communication and better security measures in comparison to 1G TDMA and CDMA, was used for multiplexing [16].

3G, on the other hand, introduced proved to fulfil promises of higher data rate and provided real broadband experience with a data rate of 2Mbps and used W-CDMA for multiplexing. Currently used 4G provide 20Mbps data rate and an IP based network [17].

4G followed by 4.5G were supposed to an efficient wireless communication system for real-time data transmission but with advents of technologies like cloud computing, IoT and mobile computing [18] the scarcity of resources raised problem, as well as these technologies, allow more open communication among nodes that raise privacy, authentication and trust issues. 5G is being developed with keeping in mind the requirement that previous technologies failed to fulfil [19]. Figure 1 illustrate the network management in 5G based IoT Architecture.

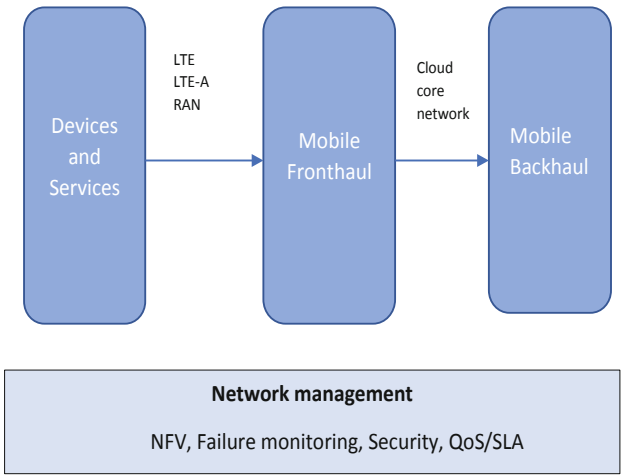

**Fig. 1.** 5G based IoT architecture

Table 1. 4G vs 5G

|  | 4G | 5G |
|---|---|---|
| Latency | 10 ms | <1 ms |
| Data traffic | 7.2 Exabytes/month | 50 Exabytes/month |
| Peak data rate | 1Gbps | 20Gbps |
| Available spectrum | 2–8 GHz | 3–300 GHz |
| Connection density | 1000/km$^2$ | 1 million/km$^2$ |
| Battery consumption | High | 10% more efficient |

## 3  5G Architectural Vision

The next-generation wireless architecture will bring in a wide range of functions and services because of the diverse quality of service (QoS) requirements as well as the proliferation of connected devices [15]. 5G is designed to fulfil the needs for highly mobile and low latency networks with more resources as shown in Table 1. Hence the architecture of 5G depicted in Fig. 2 is sliced in 4 various layers by keeping in mind the diverse requirement of Human to Human and Human to Machine connections. Layers are named as the network layer, controller layer, management and orchestration layer, and service layer [20].

5G architecture will provide dedicated services to various sectors. This on-demand network slicing will enable the automated SLA allocation for sliced instances [21]. Hence the first layer of 5G architecture is a customer-oriented layer for personalized services including augmented reality as well as personalized internet services [22].

The second layer is the most important layer of 5G network as it enables cognitive procedures implementation throughout the slice lifecycle. This layer is further divided into two sub-layers network function virtualization (NFV) and software-defined network (SDN) [23]. The heterogeneity in the network will be enabled because of automatic instantiation and allocation of resources based on needs of service. Based on the function of the network services will be allocated to various service providers that will be connected to a centralized core network thus achieving low latency and lower number of handoffs. Moreover, the software-defined nature of the network will enable the orchestration of resources their efficient management and a trustworthy secure mechanism in a heterogeneous network. The software-defined network will implement various security measures to enable the privacy of users because of the open nature of the network [24, 25].

Next layer of the architecture is radio access network (RAN) infrastructure. It acts as a pipeline for recursive allocation of resources in a highly dense network where resources need to be allocated repeatedly to fulfil similar customer-oriented requests within a network [26]. This recursive allocation makes 5G network a scalable network by the allocation of a certain instance to multiple entities to handle the workload and bringing in elasticity as network services; as part of the network is allocated recursively based on customer needs [4, 27].

The last layer of this architecture will constitute end-user devices including IoT enabled sensor devices, mobile devices, and wearable devices. The recursive allocation of resources will allow a sliced instance to be allocated to multiple tenants as virtual infrastructure for the deployment of services [28].

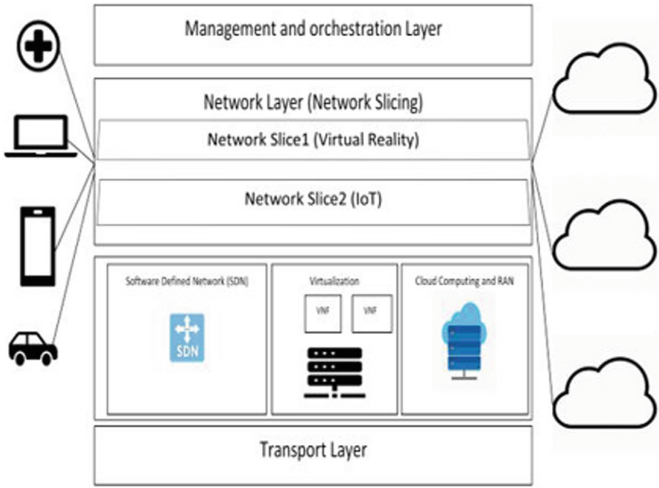

**Fig. 2.** 5G architecture

In a nutshell, 5G will be deployed on top of current 4G infrastructure but it will bring in multiple technologies like NFV and software-defined networking (SDN) [29] to handle the issues related to heterogeneity and security that openness of the 5G network will bring in. Moreover, network slicing will enable scalability and elasticity of heterogeneous network along with greater bandwidth, availability and low latency [30].

## 4 Integration of IoT and Related Technologies in Next-Generation 5G Network and Privacy

Next-generation wireless technology will be a heterogeneous network that constitutes of various services. Due to the heterogeneity and complexity 5G is divides into four various layers each providing different functionality hence possessing different security threat.

5G core architecture uses services like virtualization of resources and their assignment according to need to multiple users on a single domain. Privacy will be a major concern as third parties will be responsible for the provision of cloud services and data management. Moreover, as IoT network is a network of various types of devices [12], hence various virtualization services provider will be involved that will raise privacy and authentication issues for content and end-users might have trust issues for the service providers [22]. In a nutshell, IoT and cloud computing can be considered as two most prominent technologies in the deployment of next-generation wireless network hence possessing the privacy threats of both the technologies as well [31]. Integration of cloud into 5G is concentrated on three main cloud technologies SDN, NFV and RAN, Intelligent edge computing [32].

### 4.1 Integration of Cloud Computing in 5G

Trust, confidentiality, and privacy of content and user are most important and concerned areas of study. Moreover, privacy is a broader area as compared to the other two. Hence the integration of cloud services for virtualization in 5G will face two issues. One is in cloud user are not aware of where their data is kept or stored as user lose direct control on the content once it's in the cloud. Other is user is not aware of how their personal information is processed and in which region it is kept moreover as the users are not aware of content location, so they are also blank about what kind of privileges both public and private organizations have on the content [33].

A solution to this problem is to encrypt the content at the user end so no one except the service provider will be able to access the content. Moreover, various ciphertext decryption techniques are used some of which are symmetric, and some are based on the Bloom Filter method [34].

### 4.2 5G and SDN

5G will be a ubiquitous network of multiple technologies spread over a wide coverage area to provide a programmatic control of the whole network for the provision of services according to user needs. In SDN because of centralized control plan privacy of user and content will be at stake [35]. As there will be centralized control of the whole network pool of resources distributed denial of services attack and identity spoofing can be a major issue to mitigate resource provision to a node of a network of nodes and identity theft will be easy in SDN because of the centralized control plane. Privacy and trust are a major concern in SDN under discussion that needs to be resolved [36].

A major concern related to SDN security is Denial of service attack. SDN has a centralized controller. Controller control the entire network traffic. So, if controller is under attack then whole network faces the consequence.

Other potential threats in SDN architecture are illegal interceptor, privacy breach, fake base station, security policy conflicts. Moreover, deployment of 5G architecture over previous generation architecture so it also inherits threats from previous generations as well [37].

### 4.3 4G/5G Integration and Privacy Issues

5G will be deployed on top of existing IP-based network it will inherit privacy threats present in the 4G network along with its perks. Attacks present is 4G that 5G will inherit presented as man-in-the-middle attack, spoofing, eavesdropping, impersonation, privacy violation attack, chosen ciphertext attack, tracing attack, replay attack, collaborated attack, disclosure attack, parallel session attack and masquerade attack [38].

Although the 5G will be a ubiquitous network of various technologies but because of heterogeneity in the network it faces privacy theft issue because of increased number of handoffs with change of network that will not only affect the low latency condition of 5G but also present privacy problems like identity privacy, content privacy, location privacy, user anonymity as well as conditional and forward privacy [39].

## 5 Solutions for Privacy Preservation in 5G Enabled IoT

The need for 5G raised due to the increased number of IoT devices that need to be connected constantly for a reliable connection that 5G promises to provide in the future. State of the art of IoT considers data storage and processing in cloud for the extraction of useful information. Integration of various architecture exposes 5G to various security, privacy, and trust issues that will be highlighted at the deployment of 5G if not resolved. 5G based IoT devices are prone to user privacy threats that are subdivided into three concerns. Data privacy concerns, location privacy concerns, and identity privacy concerns.

Due to the inclusion of multiple stakeholders and heterogeneity (HetNet) of network into 5G architecture leakage of user information is not a problem. Data storage or analysis companies can access user information store it for a period longer than agreed and give it to third parties to earn an enormous amount of revenue. Table 2 show some major data privacy issues and their counter-measures.

**Table 2.** Data privacy threats

| Threat | Solution |
| --- | --- |
| Leakage of user's personal information | Data protection and privacy techniques |
| Usage of data without user's consent | End-to-End encryption schemes |
| Due to Heterogeneity third party can derive user's data using data mining | End-to-End encryption schemes |
| Malicious nodes | Authentication and exclusion |
| Trusted network issue | Authorized access to nodes |

Location-based privacy threats can be considered the most concerning attacks as service providers will be able to access user's location without them knowing. Location-based services (LBS) has the capability to turn on user location information without user knowledge also they are capable to access user's real-time location information. Table 3 presents location-based threats to the 5G-IoT network.

Hence 5G will be a network with an enhanced number of handovers raising the risk of identity theft, trust, and authentication issues. A solution proposed to preserve privacy theft in the next-generation wireless network is to software-defined networking by sharing user information among trusted access points [40]. Moreover, privacy-oriented encryption techniques are also a solution to ensure privacy in 5G enabled network [41]. Beside this using proxy with RFID tags to conceal users, location can also help in preserving user's location information when eavesdroppers have high computational resources

Table 3. Location privacy threats

| Threat | Solution |
| --- | --- |
| IoT device position revealed | Anonymity-based solutions/outlier detection and database consistency monitoring |
| Tracking user movements | Data perturbation and obfuscation methods |
| Database corruption (wrong location reference) | Outlier detection and database consistency monitoring |

[42] A solution proposed to this problem is to distribute cloud data files into specific services and files, in the end, replicating them into edge devices hence not uploading the content into the cloud for the training of the model [43]. IoT also covers the paradigm of mobile edge computing where privacy protection of edges is essential [44]. Moreover, ToRPEDO, PIERCER and IMSI cracking attack are new threats to 5G enabled IoT devices. Along with previous privacy attacks present in 5G due to the integration with past technologies like IoT 5G is vulnerable to 3 new type of identity theft attacks which are presented in Table 4 along with countermeasures Literature Proposes solutions like user identity verification for Cloud services, encryption of user location for privacy, subscriber identity encryption for preservation against IMSI attacks.

Table 4. Attacks and countermeasures

| Attack | Comments | Potential target | Counter measures |
| --- | --- | --- | --- |
| ToRPEDO attack | Access user location; inject fabricated messages and enable DOS attack it is the enabler of the other two attacks | Centralized control elements | It adds noise as fake paging message [38] |
| PIERCER attack | Enable association of victims IMSI with attacker's phone it can also enable previous attacks on the device | Identity and information theft | A proposed solution is to never send IMSI in the paging message |
| IMSI cracking attack | Using IMSI brute force attack can be initiated onto a victim's device | Roaming and user equipment | To-add IMSI catcher in the path IMSI paging sniffer connected to device [33, 34] |

## 6 Conclusion

5G is the technology of future hence possessing all the previous technologies and their flaws. Firstly, as 5G will be deployed on top of existing 4G network it will entail all the privacy issues including spoofing and impersonation threat.

Furthermore, technologies like SDN, IoT, Cloud Computing are what make 5G network so the deployment of 5G will encompass all the privacy issues of these technologies as well. In a nutshell, the paper is divided into 4 various section presenting the evolution of 5G, its architecture and privacy issues processed by each of the layers individually are also presented. Along with these privacy flaws inherited by 5G because of integration with previous technologies like SDN, IoT and Cloud Computing are also discussed in detail.

Discussion in this paper is limited to the privacy aspect of the 5G hence excluding security domain. Future work can include security of 5G and flaws presented in 5G because of integration with other technologies.